\documentclass[epj]{svjour}

\usepackage{graphics}

\begin{document}

\newcommand{\sub}[1]{_\textrm{\scriptsize{#1}}}
\newcommand{\sps}[1]{^\textrm{\scriptsize{#1}}}

\title{Magnetic properties of PrCu$_2$ at high pressure}

\author{A.~Sacchetti\inst{1}, M.~Weller\inst{1}, H.~R.~Ott\inst{1}, Y.~\=Onuki\inst{2}}

\institute{Laboratorium f\"ur Festk\"orperphysik, ETH-Z\"urich, CH-8093
Z\"urich, Switzerland \and Faculty of Science, Osaka University,
Machikaneyama, Toyonaka, Osaka 560, Japan}

\date{Received: July 17, 2008 / Revised version: \today}

\abstract{
We report a study of the low-temperature high-pressure phase diagram of the intermetallic compound PrCu$_2$, by means of molecular-field calculations and $^{63,65}$Cu nuclear-quadrupole-resonance (NQR) measurements under pressure. The pressure-induced magnetically-ordered phase can be accounted for by considering the influence of the crystal electric field on the $4f$ electron orbitals of the Pr$^{3+}$ ions and by introducing a pressure-dependent exchange interaction between the corresponding local magnetic moments. Our experimental data suggest that the order in the induced antiferromagnetic phase is incommensurate. The role of magnetic fluctuations both at high and low pressures is also discussed.
}


\maketitle

\section{Introduction}
\label{SecIntro}
The intermetallic compound PrCu$_2$ exhibits several intriguing low-temperature phase-transitions, that are still only partially understood. The room-temperature crystal structure is orthorhombic ($Imma$) and can be viewed as resulting from a small distortion of the hexagonal AlB$_2$-type structure \cite{struct}. Quasi-hexagonal Pr-sheets parallel to the crystalline $ac$ plane alternate with corrugated Cu-sheets in a stack along the $b$ axis. PrCu$_2$ is a Van Vleck paramagnet but, due to a spontaneous ordering among the Pr$^{3+}$ quadrupole moments, exhibits an induced cooperative Jahn-Teller (JT) distortion at $T\sub{JT}=7.6$~K \cite{JT,Ott}. At this transition the crystal structure changes from orthorhombic to monoclinic \cite{StructJT,AFmK1,Jorge}. At temperatures below 50~K the hard and easy magnetization axes can be switched by applying an external magnetic field in the range 10-30~T, manifested in a metamagnetic transition \cite{chi,metmag}. Neutron-diffraction experiments revealed the existence of an incommensurate antiferromagnetic (AF) order of the $4f$ Pr$^{3+}$ magnetic moments below 54~mK, concomitant with the ordering of the magnetic moments of the $^{141}$Pr nuclei \cite{AFmK1,AFmK2}.

More recently, Schenck and coworkers \cite{muons} reported, on the basis of a $\mu$SR study, an unexpected onset of incommensurate AF order among Pr$^{3+}$ moments at 65~K, at variance with the results of previous magnetic susceptibility measurements \cite{chi}. Subsequent NMR-NQR data combined with \textit{ab initio} calculations indicated that this transition cannot be considered as a bulk phenomenon and instead suggest the importance of dynamical correlations between the Pr$^{3+}$ magnetic moments \cite{NQR}. It was argued that the polarization observed with $\mu$SR may be partially due to a local compression of the lattice provoked by the implanted muon. In this context, it is recalled that the application of pressures exceeding 12~kbar establishes an AF order among the local Pr$^{3+}$ moments below $T\sub{AF} \approx 9$~K \cite{AFHP1,AFHP2}. The sharp onset of this transition at relatively low pressure suggests that PrCu$_2$ is close to a threshold beyond which AF order is stabilized. The high-pressure transition cited above was observed by means of magnetic-susceptibility and resistivity measurements \cite{AFHP1,AFHP2} and thus the magnetic structure of the ordered phase is not known. It was suggested \cite{AFHP1} that the ordered magnetic moment reaches a value close to the full moment of the Pr$^{3+}$ ions ($m\sub{Pr} = 3.58$~$\mu\sub{B}$, see sect.~\ref{SecCalc}), in spite of the fact that the transition is most likely of induced-moment type \cite{IMM}.

The aim of the present paper is to provide a more detailed discussion of the magnetic structure at high-pressure and of the interactions responsible for the magnetically ordered phase. To this end we employ mean-field calculations considering the crystal electric field (CEF) effect on the $4f$ electron orbitals of the Pr$^{3+}$ ions. In addition we present and discuss the results of NQR experiments probing the $^{63,65}$Cu nuclei. Similar calculations have successfully been employed to calculate several physical properties of PrCu$_2$ \cite{chi,CEF1,CEF2}, whereas the chosen experimental method proved to be a useful tool in studying the microscopic magnetic properties of this material \cite{NQR}.

\section{Model and calculations}
\label{SecCalc}
In PrCu$_2$ the $4f$ electron orbitals of the Pr$^{3+}$ ions adopt a $J=4$ Hund's rule ground-state and the full magnetic moment is $m\sub{Pr} = g \sqrt{J(J+1)} \mu\sub{B}$, where $g=0.8$ is the Land\'e factor. The corresponding nine-fold degenerate $^3$H$_4$ multiplet is split completely by the CEF of orthorhombic symmetry. The relevant CEF hamiltonian $\mathcal{H}\sub{CEF}$ can be written as \cite{chi,CEF1,CEF2}
\begin{eqnarray}
\nonumber
& & \mathcal{H}\sub{CEF} = B_2^0 \mathcal{O}_2^0 + B_2^2 \mathcal{O}_2^2 \\
\nonumber
& & \phantom{\mathcal{H}\sub{CEF}} + B_4^0 \mathcal{O}_4^0 + B_4^2 \mathcal{O}_4^2 + B_4^4 \mathcal{O}_4^4 \\
& & \phantom{\mathcal{H}\sub{CEF}} + B_6^0 \mathcal{O}_6^0 + B_6^2 \mathcal{O}_6^2 + B_6^4 \mathcal{O}_6^4 + B_6^6 \mathcal{O}_6^6,
\end{eqnarray}
where $B_m^n$ are the CEF parameters and $\mathcal{O}_m^n$ are the Stevens operators, defined by combinations of the angular momentum operators $J_x$, $J_y$, and $J_z$ and the $x$, $y$, and $z$ coordinates are chosen along the crystalline $c$, $a$, and $b$ axis, respectively \cite{CEF2}. A hamiltonian $\mathcal{H}\sub{qq}$ represents the coupling between the electric quadrupole moments of the Pr$^{3+}$ ions. We consider only the two dominant terms and employ the mean-field approximation, such that \cite{CEF2}
\begin{equation}
\mathcal{H}\sub{qq} = - K\sub{M} \langle \mathcal{O}_2^2 \rangle \mathcal{O}_2^2 - K\sub{JT} \langle \mathcal{O}_{xy} \rangle \mathcal{O}_{xy},
\label{EqHqq}
\end{equation}
with $\mathcal{O}_{xy}=J_x^2 - J_y^2$. The brackets $\langle...\rangle$ denote a thermal average and $K\sub{M}$ and $K\sub{JT}$ are the magneto-elastic and JT coupling strengths, respectively. The first and the second term in eq.~\ref{EqHqq} regulate the metamagnetic and JT transition, respectively. Finally, in the presence of an external magnetic field $\vec{H}\sub{ext}$, a Zeeman term
\begin{equation}
\mathcal{H}\sub{Z} = - g \mu\sub{B} \vec{H}\sub{ext} \cdot \vec{J}
\end{equation}
must be considered, as well. Here $\vec{J}=(J_x,J_y,J_z)$.

Based on the model hamiltonian \cite{CEF2}
\begin{equation}
\mathcal{H}_0 = \mathcal{H}\sub{CEF} + \mathcal{H}\sub{qq} + \mathcal{H}\sub{Z},
\end{equation}
several physical quantities, such as the magnetic susceptibility and the thermal expansion coefficients were calculated and successfully related to the experimental data for PrCu$_2$. Equally satisfactory results were previously obtained with similar models \cite{chi,CEF1}.

Here we extend the above model by augmenting the hamiltonian $\mathcal{H}_0$ with an exchange interaction
\begin{equation}
\mathcal{H}\sub{ex} = g \mu\sub{B} \vec{H}\sub{ex} \cdot \vec{J}
\label{EqHex}
\end{equation}
between the dipole moments, where $\vec{H}\sub{ex}$ is the exchange-field present at the Pr site. In an attempt to describe the new magnetic phase observed at high-pressure, we use the total hamiltonian
\begin{equation}
\mathcal{H} = \mathcal{H}_0+\mathcal{H}\sub{ex} = \mathcal{H}\sub{CEF} + \mathcal{H}\sub{qq} + \mathcal{H}\sub{Z} + \mathcal{H}\sub{ex}.
\label{EqHtot}
\end{equation}
For $\mathcal{H}\sub{ex}$ we employ again the mean-field approach in the form of
\begin{equation}
\vec{H}\sub{ex} = \sum_i{A\sub{ex}^i \langle \vec{J}_i} \rangle,
\label{EqExH}
\end{equation}
where the sum runs over all Pr ions and the $A\sub{ex}^i$ are the corresponding exchange constants in units of magnetic field, here assumed to be isotropic.

Since we expect that the exchange interaction between the Pr-ions is mainly due to the rather local Ruderman-Kittel-Kasuya-Yosida (RKKY) mechanism, we consider only nearest-neighbor interactions. For simplicity, we restrict the calculation to two magnetic sublattices. In principle the number of sublattices depends on the magnetic structure and can be much larger than two, ideally infinite for an incommensurate magnetic structure. Because a large number of sublattices implies the introduction of many unknown exchange constants this would eventually lead to an untractable model. At any rate, our attempts to introduce more than two sublattices were unsuccessful and provoked instabilities in the solution and a dramatic dependence of the results on the arbitrarily assumed exchange constants. Our choice of only two sublattices is equivalent to fix the Pr-sites on a hexagonal direct lattice and to assume a magnetic structure with a propagation vector $\vec q = (1/2,1/2,0)$ in the reciprocal hexagonal lattice. In this case, each Pr ion is surrounded by 2 nearest-neighbors belonging to the same magnetic sublattice and 4 belonging to the other one. The inter-sheet interactions can be neglected. Assuming a single exchange constant $A\sub{ex}$ for nearest-neighbor interaction, eq.~\ref{EqExH} leads to
\begin{eqnarray}
\vec{H}\sub{ex}^1 = A\sub{ex}(2 \langle \vec{J}_1 \rangle + 4 \langle \vec{J}_2 \rangle)\phantom{,}\\
\vec{H}\sub{ex}^2 = A\sub{ex}(2 \langle \vec{J}_2 \rangle + 4 \langle \vec{J}_1 \rangle),
\end{eqnarray}
where the indices 1 and 2 refer to the two sublattices. Consequently eq.~\ref{EqHex} can be rewritten as
\begin{eqnarray}
\nonumber
& & \mathcal{H}\sub{ex} = g \mu\sub{B} A\sub{ex}[(2 \langle \vec{J}_1 \rangle + 4 \langle \vec{J}_2 \rangle) \cdot \vec{J}_1\\
& & \phantom{\mathcal{H}\sub{ex}} + (2 \langle \vec{J}_2 \rangle + 4 \langle \vec{J}_1 \rangle) \cdot \vec{J}_2].
\end{eqnarray}

With the above assumptions, the diagonalization of the hamiltonian (\ref{EqHtot}) provides the corresponding eigenenergies and eigenstates for each sublattice. From these the mean fields $\langle \mathcal{O}_2^2 \rangle$, $\langle \mathcal{O}_{xy} \rangle$, and $\langle \vec{J} \rangle$ are calculated in the form of thermal averages. The values so obtained are again inserted into the hamiltonian and the procedure is repeated until a self-consistent convergence is achieved. The values $B_n^m$, $K\sub{M}$, and $K\sub{JT}$ were taken from ref.~\cite{CEF2}, whereas $A\sub{ex}$ is used as a free parameter mimicking the effect of pressure.

\begin{figure}
\resizebox{\columnwidth}{!}{\includegraphics{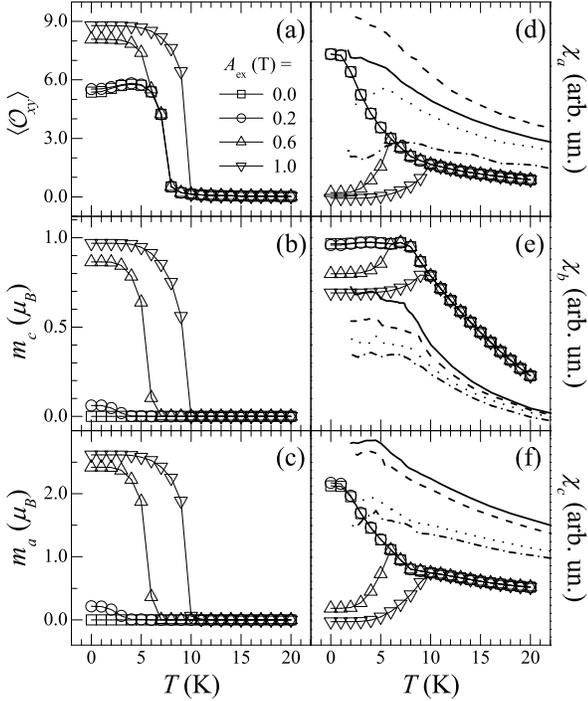}}
\caption{Temperature dependence of relevant physical quantities calculated as explained in the text for $A\sub{ex} = 0.0$~(squares), 0.2~(circles), 0.6~(up-triangles), and 1.0~T (down-triangles). (a) JT quadrupole moment $\langle \mathcal{O}_{xy} \rangle$; (b,c) magnetic moments along $c$ and $a$ axes, respectively; (d,e,f) magnetic susceptibility along the $a$, $b$, and $c$ axes, respectively. Thin solid lines are guides to the eye. Thick lines in panels (d,e,f) represent experimental data reproduced from \cite{AFHP2} at 3 (solid lines), 11 (dashed lines), 15 (dotted lines), and 18~kbar (dash-dotted lines). The comparison in panels (d,e,f) is with respect to the temperature dependence only.}
\label{FigCalc}
\end{figure}

The main results of our calculations are summarized in Fig.~\ref{FigCalc}, where the calculated temperature dependencies of several relevant physical quantities for different values of $A\sub{ex}$ are shown.
Panel (a) displays the self consistent ordered quadrupole moment $\langle \mathcal{O}_{xy} \rangle$. This quantity is non-zero only below the JT phase transition. An increasing $A\sub{ex}$ has a moderate effect on the JT-transition by slightly enhancing both $T\sub{JT}$ and the saturation value of $\langle \mathcal{O}_{xy} \rangle$. Panels (b) and (c) represent the calculated $c$ and $a$ components of the magnetic moment $\vec{m} = g \mu\sub{B} \langle\vec{J}\rangle$ for one sublattice.\footnote{In zero external field the magnetic moments on the second sublattice are oriented exactly antiparallel to those of sublattice 1 because of the AF exchange coupling.} The calculated $b$ component is always zero. It is clear that upon increasing $A\sub{ex}$, an AF order rapidly develops at low temperatures with an ordered magnetic moment in the $ac$ plane. The magnetic moment orientation forms an angle of $\approx 15^\circ$ with the $a$ axis and the saturation value for $A\sub{ex} \geq 0.6$~T is of the order of 2.8~$\mu\sub{B}$, i.e., close to the value of 3.58~$\mu\sub{B}$ of the full ionic moment of the Pr$^{3+}$ ion (see above). These results indicate that our simple model is able to reproduce the AF magnetic order that is experimentally observed at high pressures below 9~K \cite{AFHP1,AFHP2}. The agreement with the experimental data \cite{AFHP1,AFHP2} is even more convincing if we compare the calculated magnetic susceptibilities with the experimental results presented in Ref.~\cite{AFHP2}. In our model this quantity can easily be calculated by introducing a small external magnetic field and performing a numerical derivative of the magnetization
\begin{equation}
\chi_\alpha = \frac{\partial m_\alpha}{\partial H_\alpha},
\end{equation}
where $\alpha = a,b,c$ and $m_\alpha = g \mu\sub{B} \langle J_\alpha \rangle$ and $H_\alpha$ are the $\alpha$ components of the magnetic moment or the magnetic field, respectively. The magnetic susceptibilities calculated in this way are shown in Fig.~\ref{FigCalc}(d-f) and are compared with the experimental curves obtained by Naka \textit{et al.} \cite{AFHP2}. The agreement with respect to the overall shape is remarkable and supports the validity of our model for describing the high-pressure phase, at least qualitatively.

\begin{figure}
\resizebox{\columnwidth}{!}{\includegraphics{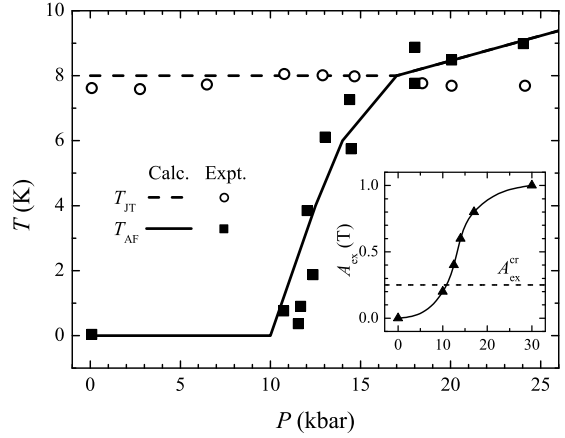}}
\caption{Pressure-temperature phase diagram of PrCu$_2$ for $P < 26$~kbar and $T<10$~K. Open and full symbols reproduce the experimentally determined values of $T\sub{JT}$ and $T\sub{AF}$, respectively (data from \cite{AFHP1,AFHP2}). Dashed and solid lines represent $T\sub{JT}$ and $T\sub{AF}$, respectively, calculated within the molecular-field CEF model, as described in the text. Inset: pressure dependence of the exchange coupling constant, $A\sub{ex}$, the free parameter in the calculation. The solid line is a guide to the eye. The dashed line represents the critical value $A\sub{ex}\sps{cr}$ (see text).}
\label{FigPD}
\end{figure}

The results of the calculations indicate that the growth of the exchange interaction between the $4f$ Pr$^{3+}$ magnetic moments can lead to the onset of a magnetically ordered phase of PrCu$_2$ at high pressures. In a second step, we attempted to model the experimental $P-T$ phase diagram. From our calculations, the values of $T\sub{JT}$ and $T\sub{AF}$ are obtained as the temperatures at which the onsets of nonzero values of $\langle \mathcal{O}_{xy} \rangle$ and $m_{a,c}$, respectively, are established. The comparison between the calculated and the experimental phase diagram is shown in Fig.~\ref{FigPD}.
The calculated phase boundaries were optimized with respect to the experimental results by adjusting the \textit{a priori} unknown pressure dependence of $A\sub{ex}$. The results of the calculations displayed in the main frame of Fig.~\ref{FigPD} are obtained with $A\sub{ex}(P)$ as shown in the inset. Although this result is to a certain extent arbitrary, it clearly shows that the $P-T$ phase diagram of PrCu$_2$ can be explained by simply introducing a pressure-dependent exchange coupling between nearest-neighbor Pr-ions. It is worth to note that the value of $A\sub{ex}$ in the region where $T\sub{AF}=0$ cannot precisely be determined and the most we can say is that it must be lower than its critical value $A\sub{ex}\sps{cr}=0.25$~T, i.e. the value above which a magnetic order is established. According to the calculation, $T\sub{JT}$ and $T\sub{AF}$ coincide above 17~kbar, in disagreement with what was claimed from experimental observations. In this respect it is worth noting that, while a quadrupole-ordered (i.e. JT) phase is possible without magnetic order, the opposite appears rather unlikely. If the quadrupole moments are ordered, the direction of the related dipole moments is fixed, but the sign of the orientation is not. On the other hand, if the dipole moments are ordered (magnetic order) the quadrupole moments must follow this order as they are tightly bound to the dipoles \cite{TAFJT}. The calculated $T\sub{JT}$ is indeed pressure independent up to 17~kbar above which it tracks the magnetic transition. Although we register this discrepancy between experiments and calculations, we also note a large spread of the $T\sub{AF}$ values obtained from the experiment \cite{AFHP1,AFHP2}. It may well be that the two transitions actually coincide above 17~kbar as suggested by our calculations and the quoted arguments of Ref.~\cite{TAFJT}.

The presented calculations provide two important results. First, the overall $P-T$ phase diagram of PrCu$_2$ can be explained by introducing a pressure dependent exchange coupling between the $4f$ Pr$^{3+}$ magnetic moments. The second result is the large ordered magnetic moment of the order of the full Pr$^{3+}$ $4f$ moment, lying in the $ac$ plane, almost parallel to the $a$ axis, and thus consistent with the previous claims based on experiments \cite{AFHP1,AFHP2}. The configuration of the magnetic structure at high pressure is still unknown, however. In particular it needs to be established whether it is commensurate or not. Due to the employed simplified assumptions for the exchange constants, this is out of reach of the present model. Despite of this, we are confident that our model sufficiently captures the magnetic properties of PrCu$_2$, since the exchange interaction between the Pr-ions is driven, as mentioned above, by the rather local RKKY mechanism. Since our model is based on a molecular-field approximation, it cannot provide any information on the role of the magnetic and quadrupolar fluctuations. The next section is devoted to some clarification of these points.

\section{NQR measurements}
\label{SecNQR}
The PrCu$_2$ powders employed in the $^{63,65}$Cu NQR experiments are from the same batch as those studied in Ref.~\cite{NQR} and were prepared as described in Ref.~\cite{sample}. The procedure and the experimental set-up for measuring NQR spectra and relaxation rates are the same as those described in \cite{NQR}. High pressures were generated in a Be:Cu piston-cylinder cell. As a pressure gauge we used Cu$_2$O for which the pressure dependence of the $^{63}$Cu-NQR signal is known \cite{Cu2O}. Both, the sample and the pressure-gauge material were powdered and embedded in paraffin, which kept them in their separate radio-frequency coils while loading the pressure cell. Silicon-oil was used as pressure transmitting medium. Due to the reduced sample-dimensions the amplitude of the measured spin-echoes is rather small and in some cases close to the detection limit of our setup.

\begin{figure}
\centering
\resizebox{0.90\columnwidth}{!}{\includegraphics{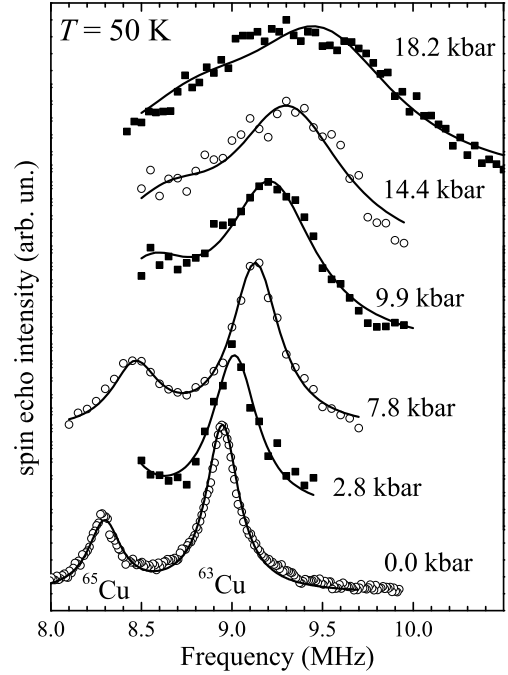}}
\caption{$^{63,65}$Cu-NQR spectra of PrCu$_2$ at 50~K and as a function of pressure. Data at ambient pressure are from \cite{NQR}. Solid lines are two-lorentzians fits to the data.}
\label{FigSpcP}
\end{figure}

\begin{figure}
\centering
\resizebox{0.90\columnwidth}{!}{\includegraphics{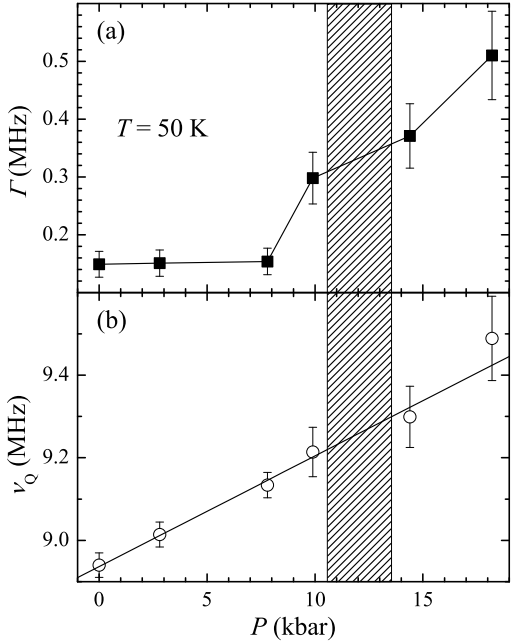}}
\caption{$^{63,65}$Cu-NQR linewidth (a) and frequency (b) of PrCu$_2$ at 50~K and as a function of pressure. Solid line in (a) is a guide to the eye, solid line in (b) is a linear fit to the data. The shaded area marks the pressure region where the magnetic transition sets in.}
\label{FigfwP}
\end{figure}

At ambient pressure, the $^{63,65}$Cu-NQR signal of PrCu$_2$ is optimal at 50~K, the temperature below which the spin-spin relaxation rate (SSRR) $T_2^{-1}$ increases significantly with decreasing temperature \cite{NQR}. Therefore the first set of NQR measurements as a function of pressure was made at 50~K and the result is shown in Fig.~\ref{FigSpcP}.
The shape of the spectra up to 7.8~kbar is very similar to that recorded at ambient pressure \cite{NQR}. A quite dramatic line-broadening is observed for $P \geq 9.9$~kbar. The spectra turn out to be so broad that the signals ascribed to the $^{65}$Cu and $^{63}$Cu nuclei, respectively, can no longer be resolved. At the same time we note a pressure induced increase of the NQR frequency. Analogous to the analysis in our previous ambient-pressure study, we fitted the spectra with two constrained lorentzian functions \cite{NQR}. From this procedure the $^{63}$Cu-NQR frequency $\nu\sub{Q}$ and linewidth $\Gamma$ can be obtained. The pressure dependence of these two quantities is shown in Fig.~\ref{FigfwP}.
The abrupt onset of line broadening starting around 10~kbar is evident in panel (a). The proximity of this anomaly to the pressure where magnetic order sets in strongly suggests that the two effects are related. The pressure-induced enhancement of the NQR-frequency $\nu\sub{Q}$ is instead monotonous, growing linearly with no significant anomaly in the explored pressure-range [Fig.~\ref{FigfwP}(b)].

Since we observe no broadening in the NQR signal of the Cu$_2$O pressure gauge, it is rather unlikely that the anomaly in $\Gamma(P)$ is due to pressure gradients inside the cell. Its origin is no doubt related to physical properties of the sample material and, in particular, it may be ascribed to a distribution of either different electric-field gradients or local magnetic fields at the Cu-site. The onset of a magnetic transition at $12\pm 2$~kbar and the corresponding anomalies observed in the pressure dependence of the magnetic susceptibility up to 70~K \cite{AFHP2} suggest that the broadening is of magnetic origin. On the other hand the smooth variation of $\nu\sub{Q}$ and the absence of any structural transition up to 40~kbar \cite{AFHP1} indicate the absence of dramatic changes of the CEF at the Cu nuclei at high pressure. At any rate, it is rather surprising that a magnetic transition with an onset around 9~K seems to be reflected in an alteration of the NQR spectrum at 50~K. It is remarkable that the line-broadening is also observed at 18.2~kbar and 100~K. This observation has, of course, no simple explanation, but we suspect that it may be ascribed, at least partially, to the presence of large magnetic fluctuations preceding this transition.

\begin{figure}
\centering
\resizebox{0.90\columnwidth}{!}{\includegraphics{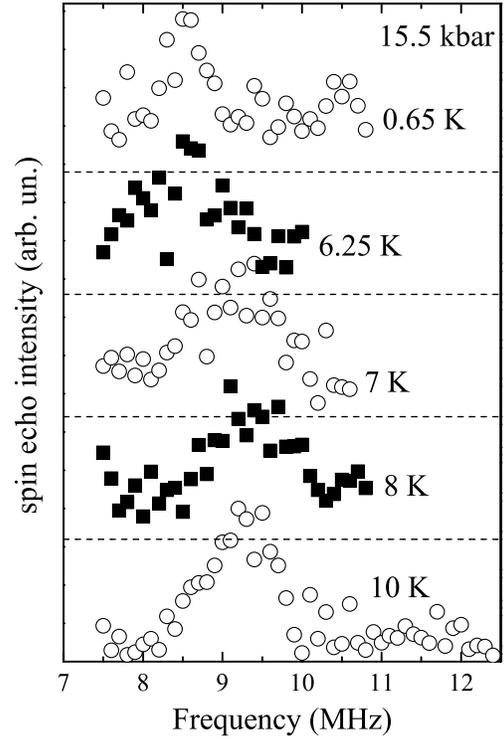}}
\caption{$^{63,65}$Cu-NQR spectra of PrCu$_2$ at 15.5~kbar and as a function of temperature. Dashed lines indicate the zero-intensity line for each spectrum.}
\label{FigSpcT}
\end{figure}

In order to gain more information on the transition itself, we repeated the NQR experiments at 7.8 and 15.5~kbar and various different temperatures between 0.65 and 300~K. The spectra at 7.8~kbar (not shown here) exhibit a temperature dependence very similar to that observed at ambient pressure \cite{NQR}. Significant differences were, however, observed at 15.5~kbar, consistent with the results shown in Fig.~\ref{FigSpcP}. The NQR spectra recorded at 15.5~kbar for $T\leq 10$~K are shown in Fig.~\ref{FigSpcT}.
On approaching $T\sub{AF} \simeq T\sub{JT}$ from above, the spin-echo intensity is progressively reduced and eventually reaches our detection limit below 6.25~K. A tiny spin-echo is then recovered only at the much lower temperature of 0.65~K. A reduction of the spin-echo intensity around $T\sub{JT}$ was also observed at ambient pressure. It is most likely due to the short SSRR measured in this region \cite{NQR}. The peak recorded at 10~K and displayed at the bottom of Fig.~\ref{FigSpcT} can be identified as the convolution of the $^{63}$Cu and $^{65}$Cu NQR signals, respectively. Upon cooling we note a blurring of this peak. The spectral weight is transferred to lower frequencies, and the signal extends over a broader spectral range. Because of the large noise in the data and of the anomalous shape of the spectra, it is not possible to extract $\nu\sub{Q}$ and $\Gamma$.

A tentative description of the results shown in Fig.~\ref{FigSpcT} can be as follows. The spectral red-shift is probably due to the JT transition as already observed at ambient pressure \cite{NQR}. Upon further cooling, the system enters the magnetically-ordered phase at $T\sub{AF} \simeq 6.5$~K (see Fig.~\ref{FigPD}) and the ordered magnetic moments of the Pr$^{3+}$ ions induce a local magnetic field at the $^{63,65}$Cu nuclei. If the order is commensurate, there will be a finite number of magnetically inequivalent Cu-sites with different local magnetic fields and consequently the NQR lines should split into multiplets. For incommensurate order, all Cu-sites are magnetically inequivalent. Due to the continuous distribution of local magnetic fields, the NQR lines simply broaden. In both cases one expects a redistribution of spectral weight over a broader spectral range and a decrease of the overall amplitude of the signal. In our data there is no evidence for the line splitting expected for a commensurate order. We argue that in the present case the above mentioned broadening is so severe that the spin-echo falls below the detection limit below 6.25~K. Upon further cooling, the ordered magnetic moment saturates and the spectral weight transfer stops. Eventually, at much lower temperatures, the large difference between the thermal populations of the nuclear levels provokes an enhancement of the spectral intensity until it exceeds the noise level again at 0.65~K. The presence of a diffuse background in the spectrum at this temperature is an additional indication for an incommensurate order, as is the AF magnetic phase at ambient pressure below 54~mK which suggests that PrCu$_2$ tends to adopt an incommensurate magnetic order \cite{AFmK1,AFmK2}. Next we argue that the line broadening due to the AF order must be expected to be extremely large. We showed in our previous work that a ferromagnetic alignment of the Pr$^{3+}$ magnetic moments produces a transferred hyperfine field at the Cu sites of 0.54~T/$\mu\sub{B}$ \cite{NQR}. Our calculation indicates the ordered magnetic moment is of the order of $m\sub{Pr}=3.58$~$\mu\sub{B}$. If these moments would order ferromagnetically, the magnetic field at the $^{63,65}$Cu nuclei would be as high as 1.9~T, which corresponds to a line splitting of 42~MHz. Of course, in an AF phase some compensation between the transferred fields produced by the staggered Pr-moments may be expected but even if this effect amounts to an order of magnitude reduction, the broadening would still be as high as 4~MHz. If this is compared with the explored spectral range (see Fig.~\ref{FigSpcT}), it is not surprising that we are not able to observe the entire broadened line in the present experiment.

\begin{figure}
\centering
\resizebox{0.90\columnwidth}{!}{\includegraphics{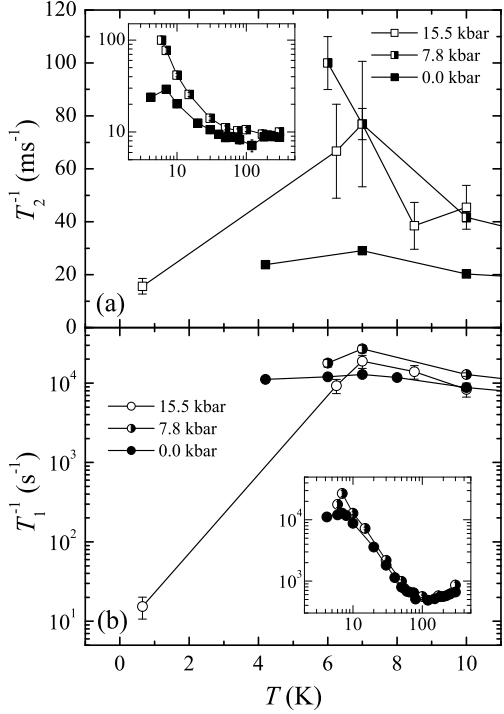}}
\caption{$^{63,65}$Cu-NQR spin-spin (a) and spin-lattice (b) relaxation rates of PrCu$_2$ as a function of temperature and for different pressures: 0.0 (full symbols), 7.8 (half-filled symbols), and 15.5~kbar (open symbols). Insets display enlarged temperature ranges. Solid lines are guides to the eye.}
\label{FigT1T2T}
\end{figure}

With the aim to gain information on the dynamics of the local magnetic moments of PrCu$_2$ at high pressure, we also measured the spin-lattice relaxation rate (SLRR) $T_1^{-1}$ and the SSRR $T_2^{-1}$, which were extracted from the time-dependencies of the transverse and longitudinal components of the nuclear magnetization, respectively. Similar to the ambient-pressure data, the transverse magnetization recovery has a gaussian shape, whereas the longitudinal component relaxes with the standard exponential time-dependence \cite{NQR}. The temperature dependencies $T_2^{-1}(T)$ and $T_1^{-1}(T)$ are shown in Fig.~\ref{FigT1T2T} for three different pressures.
It may be seen that the relaxation rates at 0 and 7.8~kbar both display a very similar temperature dependence (see insets in Fig.~\ref{FigT1T2T}). This confirms, as argued above, that pressure induces only negligible changes in the magnetic properties of PrCu$_2$ before the AF order sets in. Both the SLRR and the SSRR are slightly larger at 7.8~kbar, the difference being more pronounced for $T_2^{-1}$. This suggests that upon approaching the transition, magnetic fluctuations are present. Down to 7~K the relaxation rates measured at 15.5~kbar are of similar magnitude as those measured at lower pressure. Upon further cooling to 0.65~K, $T_2^{-1}$ decreases by a factor of approximately 4 and $T_1^{-1}$ exhibits a dramatic drop of more than three orders of magnitude.

We interpret these data as follows. On approaching $T\sub{JT}$ from above, increasing fluctuations of the electric-field gradient lead to a slight enhancement of the relaxation rates. As the system enters the quadrupole-ordered JT phase, these fluctuations are reduced and both $T_1^{-1}(T)$ and $T_2^{-1}(T)$ cross a weak maximum at $T\sub{JT}$. At low pressure ($P<12$~kbar) the magnetic moments are disordered in the JT phase and sizable magnetic fluctuations below $T\sub{JT}$ lead to rather high relaxation rates at low temperature. At 15.5~kbar, however the JT transition is accompanied by AF ordering and the magnetic degrees of freedom are reduced, as well. Consequently a much larger decrease in the relaxation rates is expected and observed. Our previous \textit{ab initio} calculations \cite{NQR} show that the Cu-NQR quantization axis lies in the $bc$ plane. According to the model calculations presented above, the ordered magnetic moments in the AF phase are oriented almost parallel to the $a$ axis, i.e. almost orthogonal to the NQR quantization axis. It is reasonable to expect a similar orientation also for an incommensurate AF order with staggered magnetic moments. Since we expect that the magnetic order predominantly affects the magnetic fluctuations along the ordered moment, it seems reasonable that in the AF phase the transverse fluctuations are more suppressed than the longitudinal ones and, as observed, $T_1^{-1}$ is much more drastically reduced than $T_2^{-1}$.

\section{Summary and conclusions}
We studied the low-temperature high-pressure phase diagram of PrCu$_2$ by means of CEF molecular-field calculations and NQR measurements. The calculations show that the phase diagram can be explained by introducing a pressure-dependent exchange interaction between the $4f$ Pr$^{3+}$ magnetic moments. The calculated ordered magnetic moment in the AF phase is almost parallel to the $a$ axis and of the order of the full Pr-moment $m\sub{Pr}=3.58$~$\mu\sub{B}$. The spectra collected in the pressure-induced AF phase suggest that the magnetic structure is incommensurate, analogous to the ambient-pressure AF magnetic order observed below 54~mK \cite{AFmK1,AFmK2}. Since the exchange coupling is most likely provoked by the RKKY mechanism, it is rather local and we expect that our model captures the main features of the magnetic structure. Therefore we propose that the staggered magnetic moment in the incommensurate phase varies with an amplitude comparable with $m\sub{Pr}$ and is almost parallel to the $a$ axis. The SLRR drops dramatically in the ordered phase and it is much smaller than that observed at ambient and low pressure. This suggests that the magnetically-disordered low-pressure phase of PrCu$_2$ is dominated by magnetic fluctuations. The pressure-induced magnetic order seems to be reflected in the NQR spectra in the form of an anomalous line-broadening at temperatures above the onset of this order. The data shown in Fig.~\ref{FigSpcP} imply that the magnetic subsystem is close to an instability over an extended range of temperatures far above $T\sub{AF}$. It is conceivable that frustration effects are responsible for this behavior because, as mentioned above, the Pr-ions occupy a sublattice with almost hexagonal symmetry.

Further studies on this material under pressure, in particular neutron diffraction experiments, would help to clarify whether the magnetic structure is commensurate or not and verify the calculated large value of the ordered magnetic moment.

\begin{acknowledgement}
We wish to thank J. Hinderer, H. R. Aeschbach, R. Monnier, J. Gavilano, R. Helfenberger, and S. Blum for their valuable contributions. We also acknowledge the financial support of the ``Schweizerische Nationalfonds f\"ur wissenschaftliche Forschung'' within the NCCR program MaNEP.
\end{acknowledgement}

\end{document}